\documentstyle[prd,aps]{revtex}

 \mathchardef\ScriptA="7241
 \mathchardef\ScriptB="7242 
 \mathchardef\ScriptC="7243
 \mathchardef\ScriptD="7244
 \mathchardef\ScriptE="7245
 \mathchardef\ScriptF="7246
 \mathchardef\ScriptG="7247
 \mathchardef\ScriptH="7248
 \mathchardef\ScriptI="7249
 \mathchardef\ScriptJ="724A
 \mathchardef\ScriptK="724B
 \mathchardef\ScriptL="724C
 \mathchardef\ScriptM="724D
 \mathchardef\ScriptN="724E
 \mathchardef\ScriptO="724F
 \mathchardef\ScriptP="7250
 \mathchardef\ScriptQ="7251
 \mathchardef\ScriptR="7252
 \mathchardef\ScriptS="7253
 \mathchardef\ScriptT="7254
 \mathchardef\ScriptU="7255
 \mathchardef\ScriptV="7256
 \mathchardef\ScriptW="7257
 \mathchardef\ScriptX="7258
 \mathchardef\ScriptY="7259
 \mathchardef\ScriptZ="725A
 \mathchardef\ScriptZ="725A

 \mathchardef\#="0023
 \mathchardef\$="0024
 \mathchardef\%="0025
 \mathchardef\ddash="705C
 
 \mathchardef\lwavy="336E
 \mathchardef\rwavy="336F
 \mathchardef\biglwavy="331A
 \mathchardef\bigrwavy="331B
 \mathchardef\bigglwavy="3328
 \mathchardef\biggrwavy="3329
 \mathchardef\littlesum="0350

\tighten
\draft
\begin{document} 
\bibliographystyle{prsty}

\preprint{TOKAI-HEP/TH-9801}
\title{
Comment on Radiative Neutrino Mass Matrix\\
with a Sterile Neutrino
}

\author{
Yutaka Okamoto$^{a}$
\footnote{E-mail:8jspd007@keyaki.cc.u-tokai.ac.jp}
and Masaki Yasu${\grave {\rm e}}^{a,b}$
\footnote{E-mail:yasue@keyaki.cc.u-tokai.ac.jp}
}

\address{\vspace{5mm}$^{a}${\sl Department of Physics, Tokai University}\\
{\sl 1117 KitaKaname, Hiratsuka, Kanagawa 259-1292, Japan}}
\address{\vspace{2mm}$^{b}${\sl Department of Natural Science\\School of Marine
Science and Technology, Tokai University}\\
{\sl 3-20-1 Orido, Shimizu, Shizuoka 424-8610, Japan}}
\date{TOKAI-HEP/TH-9801, December, 1998}
\maketitle

\begin{abstract}
A mechanism of radiatively generating neutrino masses is implemented in an 
$SU(2)_L\times U(1)_Y\times SU(2)^\prime_L\times U(1)^\prime_Y$ model, where 
the first and second families respect $SU(2)_L\times U(1)_Y$ while 
$SU(2)^\prime_L\times U(1)^\prime_Y$ is specific for the third family.  
The fourth neutrino, $\nu_s$, that has a $U(1)_Y\times U(1)^\prime_Y$ coupling 
joins in the model to induce neutrino mixings by additional interactions with 
$e$ and $\mu$.  The phenomenologically consistent oscillation of $\nu_s$ - $\nu_e$ 
requires a dominated coupling of $\nu_s$ to $e$. 
\end{abstract}


\section{Introduction}

New era of physics beyond the standard quark-lepton physics has been opened 
by the SuperKamiokande collaboration who has provided the convincing evidence on 
atmospheric neutrino oscillations.\cite{SuperKamiokande}  There have also been 
other observed anomalous phenomena suggesting that neutrinos are oscillating, 
which are often referred to the solar neutrino deficit
\cite{SolarDeficit} and to the LSND signal\cite{LSND}.  Such neutrino oscillations 
are only possible if neutrinos are massive\cite{EarlyMassive}. The well known 
theoretical frameworks realizing neutrino oscillations are based either on a seesaw 
mechanism\cite{SeeSaw,HigherDim} implying a huge mass scale such as the unification scale 
or on a radiative mechanism\cite{Radiative} employing extra Higgs bosons 
at the electroweak scale.  These observed oscillations could be explained in general by  
the simultaneous oscillations among three known neutrinos, 
$\nu_{e,\mu,\tau}$\cite{ThreeOscillation}.  However, it seems reasonable  
to assume that these phenomena really occur as a result of the oscillations involving only 
two neutrino species.  Along this line of thought, it has been suggested 
that transitions of $\nu_\mu\leftrightarrow\nu_\tau$ (atmospheric), 
$\nu_e\leftrightarrow\nu_s$ (solar) and $\nu_e\leftrightarrow\nu_\mu$ (LSND) explain 
neutrino oscillations indicated by the observed data, where $\nu_s$ is a 
sterile neutrino with no interactions in the standard model\cite{TwoOscillation}.  

In the radiative mechanism\cite{Radiative}, an extra Higgs doublet is required 
to yield a coupling to a charged Higgs singlet that interacts with a neutrino-charged 
lepton pair. The smallness of neutrino masses comes from the smallness of charged lepton 
masses and from feeble couplings associated with interactions of the neutrino-charged 
lepton pairs. The extensive analyses on neutrino physics arising from the radiative 
mechanism have been performed in detail\cite{ZeeModelNU} and have shown that 
the radiative mechanism gives consistent results with the observed data.  Therefore, 
it is of quite importance to construct a gauge model that includes a sterile neutrino.  
In the recent study done by N. Gaur {\it et al.}\cite{Sterile}, it has been explicitly 
demonstrated how the sterile neutrino scenario works well in their model with radiatively 
generated neutrino masses.  Since all neutrinos are kept 
massless at the tree level, a sterile neutrino, which is a gauge singlet in the standard 
model, should be protected from acquiring a Majorana mass.  The simplest mechanism is 
based on an extra $U(1)^\prime$ symmetry.  If the scenario really shoots the right way 
to go beyond physics of the standard model, there must exist a physical reason of the 
need for the $U(1)^\prime$ symmetry\cite{PhysicalReason}.  

In this report, the mass protection $U(1)^\prime$ symmetry is identified with another 
hypercharge in the standard model.  The extended gauge symmetry to be studied is $SU(2)_L$ 
$\times$ $U(1)_Y$ $\times$ $SU(2)^\prime_L$ $\times$ $U(1)^\prime_Y$, where the first and 
second families respect $SU(2)_L$ $\times$ $U(1)_Y$ while $SU(2)^\prime_L$ $\times$ 
$U(1)^\prime_Y$ is specific for the third  family\cite{ThirdFamilyZ,ThirdFamilyColor}.  
Since the third 
family carries different quantum numbers from those of the first and second families,  
Through mixing effects due to extra gauge bosons, $W$ and $Z$ interactions with 
the third family will differ from those with the first and second families .  
Investigation of phenomenological effects due to the non-universality specific to the third 
family is of great interest\cite{Phenomen}; however, this subject will be discussed 
in a separate article\cite{OurPhenomen}.    
One may wonder if the quark mixings with the third family are realized to occur.  In fact, 
the mixings cannot be accommodated unless a new interaction is introduced.  
A possible mechanism will be briefly discussed in the last section.   
In the present discussions, we concentrate on the study of radiative mechanism 
by simply assuming that these extra gauge bosons are heavy enough.  
In the next section, a gauge model based on 
$SU(2)_L$ $\times$ $U(1)_Y$ $\times$ $SU(2)^\prime_L$ $\times$ 
$U(1)^\prime_Y$ is formulated.  Neutrino masses are generated by  
the radiative mechanism based on one-loop diagrams, which is discussed  
in the section 3.  The last section is devoted to summary and discussions.

\section{Outline of the Model}

Our extended gauge model with an $SU(2)_L$ $\times$ $U(1)_Y$ $\times$ 
$SU(2)^\prime_L$ $\times$ $U(1)^\prime_Y$ symmetry is arranged such 
that the first and second families carry quantum numbers 
of $SU(2)_L$ $\times$ $U(1)_Y$ while the third family transforms under 
$SU(2)^\prime_L$ $\times$ $U(1)^\prime_Y$. Its spontaneous breakdown to $SU(2)$ $\times$ $U(1)$ 
will be induced by Higgs scalars whose quantum numbers are placed as 
($SU(2)_L$, $U(1)_Y$, $SU(2)^\prime_L$, $U(1)^\prime_Y$):
\begin{equation}\label{Eq:OurHiggsXi}
\xi_1 : ({\bf 2}, 0, {\bf 2}, 0),
\end{equation}
for $SU(2)_L$ $\times$ $SU(2)^\prime_L$ $\rightarrow$ $SU(2)$ by the vacuum expectation 
value (VEV) of $\langle 0|\xi_1|0 \rangle$ ($\propto$ $I$) and 
\begin{equation}\label{Eq:OurHiggsSmallPhi}
\Phi_S : ({\bf 1}, 1/2, {\bf 1}, -1/2),
\end{equation}
for $U(1)_Y$ $\times$ $U(1)^\prime_Y$ $\rightarrow$ $U(1)$ by $\langle 0|\Phi_S|0 \rangle$. 
The gauge bosons, 
$W^0_\mu$ and $B^0_\mu$, associated with $SU(2)$ $\times$ $U(1)$ are then described by 
\begin{equation}\label{Eq:EffectiveSTD}
W^0_\mu = \cos\theta_L V_\mu + \sin\theta_L V^\prime_\mu ,\hspace{3mm}
B^0_\mu = \cos\theta_Y Y_\mu + \sin\theta_Y Y^\prime_\mu ,
\end{equation}
where $V_\mu$, $Y_\mu$, $V^\prime_\mu$ and $Y^\prime_\mu$ are, respectively, 
gauge bosons of $SU(2)_L$, $U(1)_Y$, $SU(2)^\prime_L$ and $U(1)^\prime_Y$ with the 
notation of $W^0_\mu$ 
= $\sum^3_{a=1}$ $(\tau^{(a)}/2)W^{0(a)}_\mu$ and similarly for $V_\mu$ and 
$V^\prime_\mu$. Let $g_L$, $g_Y$, $g^\prime_L$ and $g^\prime_Y$ be the gauge couplings, 
then the mixing angles are defined by 
\begin{equation}\label{Eq:Mixing}
\sin\theta_L = g_L / \sqrt{g^2_L+g^{\prime 2}_L},\hspace{3mm}
\sin\theta_Y = g_Y / \sqrt{g^2_Y+g^{\prime 2}_Y},
\end{equation}
which give the gauge couplings of $SU(2)$ $\times$ $U(1)$, $g$ and $g^\prime$, as
\begin{equation}\label{Eq:CouplingSTD}
g = \cos\theta_L g_L = \sin\theta_L g^\prime_L,\hspace{3mm}
g\prime = \cos\theta_Y g_Y = \sin\theta_Y g^\prime_Y.
\end{equation}
In addition to these scalars, 
\begin{equation}\label{Eq:OurHiggsphi}
\phi_1 : ({\bf 2}, 1/2, {\bf 1}, 0), \hspace{3mm}
\eta_1 : ({\bf 1}, 0, {\bf 2}, 1/2)
\end{equation}
are responsible for generating masses of the first and second families and 
of the third family, respectively.  

Denoting extra massive gauge bosons in $SU(2)$ $\times$ $U(1)$ by 
$W^{0\prime}_\mu$ and $B^{0\prime}_\mu$:
\begin{equation}\label{Eq:GaugeBosonsSTD}
W^{0\prime}_\mu = \cos\theta_L V^\prime_\mu - \sin\theta_L V_\mu ,\hspace{3mm}
B^{0\prime}_\mu = \cos\theta_Y Y^\prime_\mu - \sin\theta_Y Y_\mu ,
\end{equation}
we find that
\begin{eqnarray}\label{Eq:CovariantDerivative}
g_LV_\mu = gW^0_\mu - G_LW^{0\prime}_\mu,\hspace{3mm}
g^\prime_LV^\prime_\mu = gW^0_\mu + G^\prime_LW^{0\prime}_\mu, 
 \nonumber \\
g_YY_\mu = g^\prime B^0_\mu - G_YB^{0\prime}_\mu,\hspace{3mm}
g^\prime_YY^\prime_\mu = g^\prime B^0_\mu + G^\prime_Y B^{0\prime}_\mu,
\end{eqnarray}
where
$G_L$ = $g_L\sin\theta_L$, 
$G^\prime_L$ = $g^\prime_L\cos\theta_L$, 
$G_Y$ = $g_Y\sin\theta_Y$ and 
$G^\prime_Y$ = $g^\prime_Y\cos\theta_Y$.
The first and second families couple to $V_\mu$ and $Y_\mu$ while the third 
family, to $V^\prime_\mu$ and $Y^\prime_\mu$.  The universality of the $W^0$ 
and $B^0$ couplings are ensured by $SU(2)$ $\times$ $U(1)$ as expected.
The $W^\prime$ and $B^\prime$ masses, $m^{(0)}_{W^\prime}$ and $m^{(0)}_{B^\prime}$ 
are given by 
\begin{equation}\label{Eq:MassiveGaugeBoson}
m^{(0)}_{W^\prime} = \sqrt{g^2_L + g^{\prime 2}_L}v_{\xi_1}/2,\hspace{3mm} 
m^{(0)}_{B^\prime} = \sqrt{g^2_Y + g^{\prime 2}_Y}v_s/2, 
\end{equation}
where 
$v_{\xi_1}$ is a VEV defined by $\langle 0|\xi_1|0 \rangle $ = 
($v_{\xi_1}$/$2\sqrt{2}$)diag.(1, 1) and $v_s$, by $\langle 0|\Phi_S|0 \rangle$ = $v_s$/$\sqrt{2}$. 
After spontaneous breaking due to  
$\phi$ (and $\eta$), $W^0$ and $B^0$ will mix with $W^{0\prime}$ and $B^{0\prime}$ to produce 
massive $W$ and $Z$ bosons and extra $W^\prime$, $Z^\prime$ and $Z^{\prime\prime}$ bosons, 
where $(W^{0(3)}$, $B^0$, $W^{0\prime (3)}$, $B^{0\prime})$ $\rightarrow$ $(Z$, $\gamma$, 
$Z^\prime$, $Z^{\prime\prime})$.  Our sterile neutrino to be defined in the next section 
will only couple to $B^{0\prime}$, thus mainly to $Z^{\prime\prime}$.  
In case that $v_{\xi_1}$ $>>$ $v_s$, the model is approximated to be described by 
$SU(2)$ $\times$ $U(1)_Y$ $\times$ $U(1)^\prime_Y$.

\section{Neutrino Masses}

In the original Zee's radiative mechanism\cite{Radiative}, an extra Higgs 
doublet and charged singlet are required. In the present context, interactions with the following 
extra Higgs bosons: 
\begin{eqnarray}\label{Eq:ChrgedHiggs}
& \phi_2 : ({\bf 2}, 0, {\bf 1}, 1/2),\hspace{3mm}
\eta_2 : ({\bf 1}, 1/2, {\bf 2}, 0), 
\nonumber \\
& \xi_2:  ({\bf 2}, 1/2, {\bf 2}, 1/2),\hspace{3mm}
\chi^+_1:  ({\bf 1}, 1, {\bf 1}, 0), \hspace{3mm}
\chi^+_2:  ({\bf 1}, 1/2, {\bf 1}, 1/2),
\end{eqnarray}
will generate the mixing of $\nu_e$ - $\nu_\mu$.  New interactions, which are of course 
invariant under $SU(2)_L$ $\times$ $U(1)_Y$ $\times$ $SU(2)^\prime_L$ $\times$ $U(1)^\prime_Y$, 
are assumed to be further subject to the conservation of a discrete $Z_2$ symmetry as a 
$\tau$-parity that is negative for the third family, $\xi_1$ and $\xi_2$. All others have the 
positive parity.  The direct coupling of $\phi_2$ and $\eta_2$ to quarks and leptons does not 
respect the gauge invariance; therefore, it is forbidden.  
Also forbidden is the other possible interaction, $\eta^T_i\xi^\dagger_2\phi_i$ ($i$ = 1,2), 
that will give $\langle\xi_2\rangle$ $\neq$ 0 
by a tadpole coupling generated after the spontaneous breaking.  By choosing the mass squared for 
$\xi_2$ to be positive in the Higgs potential, one can safely set 
$\langle 0|\xi_2|0 \rangle$ $=$ 0 that would be disturbed if the tadpole interaction were active. 
It also ensures the absence of the tree-level 
$\nu^T_{e,\mu}\nu_\tau$ term arising from $L^{(e,\mu )T}\xi_2L^{(\tau )}$, 
where $L^{(i)}$ denote lepton doublets in three families ($i$ = $e$, $\mu$, $\tau$). 

The relevant part of the lagrangian yielding $\nu_e$ - $\nu_\mu$, is given by
\begin{eqnarray}\label{Eq:NUeNUmu}
\Delta{\cal L}_{\nu_e -\nu_\mu}  &=&  
f_{e\mu}L^{(e)T}Ci\tau^{(2)}L^{(\mu )}\chi^+_1 
+ \sum_{i=e,\mu} f_{i\tau}L^{(i)T}Ci\tau^{(2)}\xi_2L^{(\tau )}
\nonumber \\
& & + \mu (\phi^T_2i\tau^{(2)}\phi_1 + \eta^T_2i\tau^{(2)}\eta_1 )\chi^-_2 
+ \mu^\prime\chi^+_2\chi^-_1\Phi_S + H.c., 
\end{eqnarray}
where $i\tau^{(2)}$ ($C$) denotes the charge conjugation operator in $SU(2)$ (the Lorentz space). 
Similarly, an additional 
interaction given by 
\begin{eqnarray}\label{Eq:NUeNUtau}
\Delta{\cal L}_{\nu_\tau -\nu_\ell}  &=&  \mu^{\prime\prime} {\rm Tr}(\xi_2^\dagger \xi_1)\chi^+_2 
+ H.c. 
\end{eqnarray}
induces $\nu_\tau$ - $\nu_{\ell(=e,\mu)}$. The corresponding one-loop diagrams for $\nu_\ell$ - 
$\nu_{\ell^\prime}$ ($\ell$, $\ell^\prime$ = $e$, $\mu$, $\tau$) with $\ell$ $\neq$ $\ell^\prime$ 
are depicted in Fig.1.  

To give a desirable neutrino mass 
matrix as shown in 
\cite{Sterile}, 
one sterile neutrino, $\nu_s$, together with 
other two neutrals, $N_{1,2}$, joins in the model:
\begin{equation}\label{Eq:ChargedHiggsTau}
\nu_s:  ({\bf 1}, -1/2, {\bf 1}, 1/2),\hspace{3mm}
N_1: ({\bf 1}, 1/2, {\bf 1}, -1/2),\hspace{3mm}
N_2: ({\bf 1}, 0, {\bf 1}, 0).
\end{equation}
The $\tau$-parity is positive for $\nu_s$ and negative for $N_{1,2}$.  
The mass term of $N_2N_2$ is allowed but $\nu_sN_1$ is forbidden.  The extra neutrals, $N_{1,2}$, 
can also get massive by $N_1\Phi^\dagger_SN_2$. The mixing of $\nu_s$ with $\nu_{e,\mu}$ is made 
possible by adding $\Delta{\cal L}_{\nu_s -\nu_\ell}$:
\begin{eqnarray}\label{Eq:NUeNUS}
\Delta{\cal L}_{\nu_s -\nu_\ell}  &=& \sum_{i=e,\mu}h_i {\overline {\ell^{(i)}_R}}\nu_s\chi^-_2 
+ H.c., 
\end{eqnarray}
where $\ell^{(i)}$ = $e$ ($i=e$) and $\mu$ ($i=\mu$).  Shown in Fig.2 are the one-loop diagrams 
for $\nu_s$ - $\nu_\ell$. It should be noted that ${\overline {\tau_R}}\nu_s\chi^-_2$ is 
forbidden by the $\tau$ - parity.  
Our sterile neutrino interacts with $B^{0\prime}_\mu$ since it has a 
$U(1)_Y\times U(1)^\prime_Y$ coupling.  
The possible mixing of $B^{0\prime}_\mu$ 
with the $Z$ boson induces a $Z$ - $\nu_s$ coupling (also by the $\nu_s$ mixings with 
$\nu_{e,\mu ,\tau}$), which should be greatly suppressed.  The 
suppression can be realized by setting the mass of $B^{0\prime}_\mu$ to be much heavier than 
the $Z$ - mass, say, of the order of 1 TeV.  The remaining symmetry, 
$SU(2)_L$ $\times$ $SU(2)^\prime_L$ $\times$ $U(1)$, is blind for $\nu_s$.

These additional interactions are invariant under an accidental global $U(1)_\ell$ transformation.  
The nonvanishing $U(1)_\ell$ - charges ($Q_\ell$) are given by 
 1 for leptons, $-1$ for $\nu_s$ and $-2$ for $\phi_2$, $\eta_2$, $\xi_2$ and 
$\chi^+_{1,2}$, which can be regarded as an extended lepton number.  This $U(1)_\ell$ will 
be broken by $\langle\phi_2\rangle$ and  $\langle\eta_2\rangle$.  
However, since $U(1)_\ell$ is broken by VEV's with 
$\vert Q_\ell\vert$ = 2, there 
is still the conservation due to a $Z_2$ symmetry of $\exp(i\pi Q_\ell )$ as 
the $\ell$-parity\cite{Sterile} that 
is negative for leptons ($Q_\ell$ = 1) and $\nu_s$ ($Q_\ell$ = $-1$). Thus, the mixings of 
$N_{1,2}$ with leptons and $\nu_s$ are totally forbidden.  
The spontaneous breakdown generates a massless Nambu-Goldstone 
boson, which becomes massive by an explicit breaking of $U(1)_\ell$ served by, for instance, 
($\phi^\dagger_2\phi_1$  + $\eta^\dagger_2\eta_1$)$\Phi^\dagger_S$\cite{Sterile}. 

The resulting neutrino mass matrix turns out to be in the form of
\begin{eqnarray}\label{Eq:NUmass}
\left[ \begin{array}{cccc}
	0&	a&	b&	d\\
	a&	0&	c&	e\\
	b&	c&	0&	f\\
	d&	e&	f&	0
\end{array} \right],
\end{eqnarray}
where $a$, $b$ and $c$ come from Fig.1 and $d$, $e$ and $f$ from Fig.2.  The masses are 
computed to be
\begin{eqnarray}\label{Eq:MatrixEntry1}
a  &=& f_{e\mu}\mu\mu^\prime
\bigg( 
m^2_\mu - m^2_e 
\bigg) 
\frac{v_{\phi_2}}{v_{\phi_1}} G(m^2_{\phi^+_1},m^2_{\chi^+_1})
\langle 0|\Phi_S|0 \rangle , \\
b  &=& f_{e\tau}\mu\mu^{\prime\prime}
\bigg[ 
m^2_\tau\frac{v_{\eta_2}}{v_{\eta_1}}G(m^2_{\eta^+_1},m^2_{\xi^+_2}) - 
m^2_e   \frac{v_{\phi_2}}{v_{\phi_1}}   G(m^2_{\phi^+_1},m^2_{\xi^+_2})
\bigg] \langle 0|\xi^0_1|0 \rangle , \\
c  &=& f_{\mu\tau}\mu\mu^{\prime\prime}
\bigg[ 
m^2_\tau\frac{v_{\eta_2}}{v_{\eta_1}}G(m^2_{\eta^+_1},m^2_{\xi^+_2}) - 
m^2_\mu   \frac{v_{\phi_2}}{v_{\phi_1}}   G(m^2_{\phi^+_1},m^2_{\xi^+_2})
\bigg] \langle 0|\xi^0_1|0 \rangle ,
\end{eqnarray}
where $m_{e,\mu,\tau}$ represent the masses of charged leptons and 
\begin{equation}\label{Eq:G_x_y}
G(x, y) = \frac{1}{16\pi^2}\frac{1}{x-y}
\bigg[ 
\frac{\log x-\log y}{x-y}-\frac{1}{x}
\bigg].
\end{equation}
The explicit form of $G$ is subject to the assumption that 
$m_{\chi^+_1}$ $\sim$ $m_{\chi^+_2}$ $\sim$ $m_{\xi^+_2}$ $\sim$ $m_{{\bar \xi}^+_2}$.  
The remaining entries are given by
\begin{eqnarray}\label{Eq:MatrixEntry2}
d  &=& f_{e\mu}\mu^\prime m_\mu h_\mu F(m^2_{\chi^+_2},m^2_{\chi^+_1})
\langle 0|\Phi_S|0 \rangle , \\
e  &=& -f_{e\mu}\mu^\prime m_eh_e F(m^2_{\chi^+_2},m^2_{\chi^+_1})
\langle 0|\Phi_S|0 \rangle , \\
f  &=& -\sum_{i=e,\mu}f_{i\tau}\mu^{\prime\prime} m_ih_i F(m^2_{\chi^+_2},m^2_{\xi^+_2})
\langle 0|\xi^0_1|0 \rangle ,
\end{eqnarray}
where
\begin{equation}\label{Eq:F_x_y}
F(x, y) = \frac{1}{16\pi^2}
\frac{\log x-\log y}{x-y}.
\end{equation}
It should be noted that the mixings of $\nu_s$ are 
controlled by either $m_e$ or $m_\mu$ but not by $m_\tau$ because of the absence of 
${\overline {\tau_R}}\nu_s\chi^-_2$ ensured by the $\tau$ - parity.  It provides more  
suppression than what would be expected\cite{Sterile} owing to $m_{e,\mu}$ $<<$ $m_\tau$. 

An example of getting phenomenologically acceptable neutrino masses and mixings is given 
by adjusting various parameters such that 
$\vert c\vert$ $>>$ $\vert b,e \vert$ $>>$ $\vert a \vert$, $\vert c \vert$ $>>$ $\vert d,f \vert$,  
$\vert f\vert$ $>>$ $\vert b \vert$,  
and $\vert ef \vert$ $>>$ $\vert cd \vert$ $>>$ $\vert ab \vert$ 
are satisfied\cite{Simplest3Families}.  The resulting neutrino 
mass spectrum consists of almost degenerate two massive neutrinos and two extremely light ones: 
($-2ab/c$, $c$ + ($ab$ + $ef$)/$c$,  $-c$ + ($ab$ + $ef$)/$c$,  $-2ef$/$c$).  
The parameter setting that reproduces current neutrino-oscillation data is supplied by 
\begin{eqnarray}\label{Eq:Parameters}
\vert a\vert &\sim& 10^{-5}~{\rm eV},\hspace{3mm}
\vert b\vert \sim 10^{-2}~{\rm eV},\hspace{3mm}
\vert c\vert \sim 1~{\rm eV},
\nonumber \\
\vert d\vert &\sim& 10^{-4}~{\rm eV},\hspace{3mm}
\vert e\vert \sim 10^{-2}~{\rm eV},\hspace{3mm}
\vert f\vert \sim  10^{-1}~{\rm eV},
\end{eqnarray}
which give
\begin{eqnarray}\label{Eq:Parameters1}
\vert f_{e\mu} \vert / 16\pi^2  &\sim& 10^{-8},\hspace{3mm}
\vert f_{e\tau} \vert / 16\pi^2  \sim 10^{-7},\hspace{3mm}
\vert f_{\mu\tau} \vert / 16\pi^2  \sim 10^{-5},
\nonumber \\
\vert h_e \vert  &\sim& 1,\hspace{3mm}
\vert h_\mu\vert  \sim 10^{-4},\hspace{3mm}
\mu\sim M_0/10,
\end{eqnarray}
where, for simplicity, $v_{\phi_1}$ ($v_{\eta_1}$) = $v_{\phi_2}$ ($v_{\eta_2}$) is taken and 
the mass parameters, $\mu^\prime$, $\mu^{\prime\prime}$, charged scalar 
masses appearing in $G(x,y)$ and $F(x,y)$ and VEV's of $\xi_1$ and $\Phi_S$, are all set equal 
to $M_0$ presumably of the order of 1 TeV.  These numerical values are so chosen that 
\begin{equation}\label{Eq:NeutrinoMasses}
m_{\nu_e} \sim 2\times10^{-7} {\rm eV}, \hspace{3mm}
m_{\nu_s} \sim 2\times10^{-3} {\rm eV},  \hspace{3mm}
m_{\nu_\mu} \sim m_{\nu_\tau} \sim 1 {\rm eV},
\end{equation}
are realized. An appropriate mixing of $\nu_s$ and $\nu_e$ is found to be controlled by the 
dominated coupling of $\nu_s$ to $e$,  {\it i.e.} $\vert h_e\vert$ $\sim$ 1.  
For $\nu_\mu$ $\leftrightarrow$ $\nu_\tau$, they are maximally mixed with each other by its 
squared mass difference $\sim$ $4(ab$ $+$ $ef)$ $\sim$ $4\times 10^{-3}$ (eV$^2$).   
The remaining mixing angles to be denoted by $\theta$  for $\nu_e$ $\leftrightarrow$ $\nu_s$ and 
$\nu_e$ $\leftrightarrow$ $\nu_\mu$, respectively, are given by,  $\sim$ $\vert cd$ $-$ $(be$
 $+$ $af)\vert/\vert ef\vert$
and  $\sim$ $\vert b/c\vert$, which can be cast into the right 
order of the magnitudes roughly controlled by $\sin^22\theta$ $\sim$ $10^{-3}-10^{-2}$, 
as have been discussed in 
Ref.\cite{Sterile} 
for $\nu_e$ $\leftrightarrow$ $\nu_s$).

\section{Summary and Discussions}

Summarizing our discussions, we have formulated a gauge model based on 
$SU(2)_L$ $\times U(1)_Y$ $\times$ $SU(2)^\prime_L$ $\times$ $U(1)^\prime_Y$ 
in order to equip with the radiative mass generation mechanism for 
neutrino oscillations.  To make this mechanism effective, one needs extra Higgs scalars, 
especially, $SU(2)$ - singlet charged ones that connect neutrinos with the charged 
leptons.  The interactions are requested to be subject to the conservation of the $\tau$ 
- parity and of the global $U(1)_\ell$ symmetry.  As a result, 
$\nu_s$ mixes with $\nu_{e,\mu,\tau}$ via $e$ and $\mu$ only.  
The spontaneously broken $U(1)_\ell$ symmetry is replaced by the remnant 
$Z_2$ symmetry, which forbids the mixings of $\nu_{e,\mu,\tau,s}$ with 
the massive $N_{1,2}$.  Our scenario implies a rather 
large coupling of $\nu_s$ to $e$, $h^2_e$ $\sim$ 3$g^2$ ($\sim$ 1), that 
favors a consistent $\nu_s$ - $\nu_e$ mixing with 
an expected solution of the solar neutrino problem.  Therefore, among the 
$\nu_s$ - $e$ transitions supposedly described by the exchanges of bosons whose 
masses are $\ScriptO$(1 TeV), one expects that the $\chi^+_2$ - exchange gives dominant 
contributions, which alter an effective 
number of neutrino species, $N_\nu$, in the big bang nucleosynthesis. The strength is roughly 
given by $h^2_e/m^2_{\chi^+}$ corresponding to the decoupling temperature of 
$\ScriptO$(100) MeV for $m_{\chi^+}$ $\sim$ 1 TeV and $\vert h_e\vert$ $\sim$ 1.  It will add 
$\ScriptO$(0.1) to $N_\nu$, which is allowed to be larger than 3 by the recent 
refined analyses\cite{BigBangNu}. 

The specific feature of our model is that the third family is 
placed in $SU(2)^\prime_L$ $\times$ $U(1)^\prime_Y$ while the first and second families are in 
$SU(2)_L$ $\times$ $U(1)_Y$.  Because of this feature, the $b$-mixing is not accommodated.  The $b$-mixing 
can be generated, for example, by introducing a vectorlike down quark, $b^\prime$, as well as extra 
Higgs scalars, $\phi_b$, $\eta_b$ and $\Phi_b$, whose quantum numbers are taken as 
$b_{L,R}^\prime$:({\bf 1}, -1/6, {\bf 1}, -1/6), 
$\phi_b$:({\bf 2}, 1/3, {\bf 1}, 1/6),
$\eta_b$:({\bf 1}, 1/6, {\bf 2}, 1/3) and  
$\Phi_b$:({\bf 1}, 1/6, {\bf 1}, -1/6).
The interactions are given by
\begin{equation}\label{Eq:IntExtra_b}
 \sum_{i=e,\mu}
\bigg(
{\overline {b^\prime_R}} f_{b^\prime i}\phi^\dagger_bQ^{(i)}_L 
+ f_{ib^\prime}{\overline {q^{(i)}_R}}\Phi^\dagger_bb^\prime_L
\bigg) 
+ f_{b^\prime \tau}{\overline {b^\prime_R}} \eta^\dagger_bQ^{(\tau )}_L + f_{\tau b^\prime}{\overline {b_R}}\Phi_bb^\prime_L 
+ M_{b^\prime}{\overline {b^\prime}}b^\prime,
\end{equation}
where $Q^{(i)}_L$ ($q^{(i)}_R$) denote quark doublets (singlets) in three families ($i$=$e$, $\mu$, $\tau$), 
$f_{ib^\prime, b^\prime i}$ are couplings 
and $M_{b^\prime}$ is a mass of $b^\prime$. The mixing of $b$ with $d$ and $s$ is induced via 
$b^\prime$ by the seesaw-like mechanism for a dominated $M_{b^\prime}$ in mass scales of down quarks.  

Phenomenologically interested is to estimate effects from extra particles other than those in 
the standard model.  Some of them cause dangerous flavor - changing interactions that must be 
greatly suppressed\cite{ZeeModelNU}.  The suppression of $Z$ - $\nu_s$ will impose the severe 
constraint on the mass of $Z^{\prime\prime}$, which mainly arises from the gauge 
bosons associated with $U(1)_Y$ $\times$ $U(1)^\prime_Y$. The phenomenology of the extra weak bosons\cite{ExtraWZ} 
to be characterized by $\ddash$anomalies" in $t$-, $b$- and $\tau$-interactions 
will be discussed elsewhere\cite{OurPhenomen}. 


\end{document}